\documentclass[12pt,preprint]{aastex}

\def\kms{\ifmmode {\rm km\,s}^{-1} \else km\,s$^{-1}$\fi}

\def\simlt{\lower.5ex\hbox{$\; \buildrel < \over \sim \;$}}

\begin{document}

\title{Multiband VLA Observations of the Faint Radio Core of 3CR 68.1}
\author{Michael S. Brotherton\altaffilmark{1}, Chun Ly\altaffilmark{2}, 
Beverley J. Wills\altaffilmark{3}, Sally A. Laurent-Muehleisen\altaffilmark{4}, 
Wil~van~Breugel\altaffilmark{5}, R. R. J. Antonucci\altaffilmark{6}}

\altaffiltext{1}{Kitt Peak National Observatory, National Optical Astronomy 
Observatories, 950 North Cherry Avenue, P. O. Box 26732, Tucson, AZ 85726; 
mbrother@noao.edu.  The National Optical Astronomy Observatories are 
operated by the Association of Universities for Research in Astronomy, Inc., 
under cooperative agreement with the National Science Foundation.} 
\altaffiltext{2}{Steward Observatory, University of Arizona, 933 North Cherry
Avenue, Tucson, AZ 85721-0065}
\altaffiltext{3}{McDonald Observatory and Astronomy Department, University of Texas, Austin, TX 78712}
\altaffiltext{4}{Department of Physics, University of California, Davis, 
Davis, CA 95616}
\altaffiltext{5}{Institute of Geophysics and Planetary Physics, Lawrence Livermore National Laboratory, 7000 East Avenue, P.O. Box 808, L413, Livermore, CA 
94550}
\altaffiltext{6}{Physics Department, University of California, Santa Barbara,
CA 93106}
\begin{abstract}

The large, powerful Fanaroff-Riley class II radio source 3CR 68.1 is an
optically red quasar with strong evidence for dust reddening, intrinsic
ultraviolet line absorption, and X-ray absorption.  In the context of its
large, extended radio structure and the evidence for intrinsic material along 
the line of sight, it is a good candidate for a quasar seen through the edge 
of an obscuring torus proposed by unified schemes.  The compact radio core
coincident with the optical position of the quasar is extremely weak compared 
to the very bright and luminous radio lobes, suggesting that the radio core
might also suffer absorption.  We observed 3CR 68.1 at six frequencies and 
show that the radio core has a flat spectrum with no evidence for free-free 
absorption or other strong absorption mechanisms.  This result helps 
establish an empirical intrinsic lower range for core-to-lobe radio flux 
in the most powerful radio sources.

\end{abstract}



\section{Introduction}

The extreme properties of 3CR~68.1 indicate it is seen in an edge-on 
geometry (Brotherton et al. 1998), along a line of sight through dusty,
ionized gas perhaps part of an obscuring torus (e.g., Antonucci 1993).  
Such a structure seen in such a geometry helps to explain the 
dust-reddened optical continuum (A$_V$ = 1.7 mag, Simpson \& Rawlings 2000),
scattered light leading to a rising continuum polarization up to 13\% at 
2000 \AA\ rest-frame (Brotherton et al. 1998), intrinsic ultraviolet 
absorption lines including metastable He I* (Brotherton et al. 1998), 
extremely weak and probably absorbed X-ray emission (Bregman et al. 1985), 
and the large radio lobe separation (53\arcsec\ at z=1.228) and small radio core 
dominance (Bridle et al. 1994) that mark an edge-on radio structure
(e.g. Orr \& Browne 1982).  

The radio core of 3CR 68.1 is an order of magnitude fainter than that of any 
other 3CR quasar (1.1 mJy at 5 GHz, Bridle et al. 1984) and it is weak even 
among Fanaroff-Riley type II (FR II; Fanaroff \& Riley 1974) radio galaxies. 
At the same time, the radio lobes make 3CR 68.1 one of the three most 
powerful 3CR quasars at 178 MHz in the compilation by Laing et al. (1983).  
We hypothesized that the radio core might be so faint because it suffers 
free-free absorption from the obscuring material along the line of sight,
measurements of which would provide an additional probe of its physical 
conditions.  

We obtained multi-band observations of 3CR~68.1 using the 
NRAO\footnote{The National Radio Astronomy Observatory is a facility of 
the National Science Foundation operated under cooperative agreement by 
Associated Universities, Inc.} Very Large Array (VLA) in order to
test this hypothesis.  Following sections of this paper discuss the
observations and data reduction (\S 2), measurements of the core (\S 3),
our null result and its implications (\S 4).  The final section (\S 5)
summarizes our conclusions.

\section{Observations and Data Reduction}

We observed 3CR~68.1 with the VLA in its A configuration on May 24, 1998 
(program AW482), using six frequencies: 325 MHz (90 cm, P band),
1.4 GHz (20 cm, L band), 4.8 GHz (6 cm, C band), 8.4 GHz (3.6 cm, X band), 
15 GHz (2.0 cm, U band), and 22 GHz (1.3 cm, K band).  
Observations at each frequency were interleaved
to provide appropriate sampling of the uv-plane.  We observed 3C 48
as the primary flux calibrator and several phase calibrators between 
observations of 3CR~68.1.  Table 1 gives the observing log showing
the total observing time per source at each frequency. 

Calibrations were done in AIPS and in general followed standard recipes in 
the AIPS Cookbook\footnote{http://www.aoc.nrao.edu/aips/cook.html}.  
Flux calibration at 15 GHz for 3C 48 required small
modification in the uv-range compared with suggested values.
For 22 GHz and 8.4 GHz, we calibrated with clean component models using the
task CALIB.  We have not performed a polarization calibration
(the radio core is unpolarized, Bridle et al. 1994).
Following calibration we used MAPIT to self-calibrate and produce 
CLEAN maps of our source.  Figures 1-3 show these images as contour maps.
Figure 1 shows only the central core region along with the beam, as the lobes are 
weak/resolved out at these high frequencies.

\section{Measurements and the Radio Core Spectrum}

The optical position of 3CR 68.1 is Right Ascension 02H 32M 28.9S, Declination 
+34\arcdeg\ 23\arcmin\ 47\arcsec\ (J2000).  We detect a radio point source at
that position, the radio core, in all our total intensity maps except 
at 325 MHz.  We used the AIPS task JMFIT to measure the position and flux density
of the radio core.  Table 2 provides these measurements.  
The RMS values in the table are 
from JMFIT.  The uncertainties on the flux were computed by adding in quadrature the 
RMS noise in the map, plus the reported error in the fit (from JMFIT) and an assumed
3.33\% error (= 10
The uninteresting 325 MHz limit comes from running IMSTAT to determine 
the peak flux in a small box around the core position; a more conservative limit
is five times the RMS noise, or 65 mJY.  The actual 325 MHz flux of the core is
likely about 1 mJy or less.  Figure 4 plots these peak flux densities against 
frequency to show the radio spectrum of the core.

The most accurate position for the radio core measured to date comes from our
high-frequency maps and is 02H 32M 28.88S, +34\arcdeg\ 23\arcmin\ 46.7\arcsec\ (J2000).

\section{Results \& Discussion}

We have detected the radio core of 3CR 68.1, coincident with the 
optical position, at 5 of 6 frequencies.  Our total intensity map
at 4.8 GHz (Fig. 2, right) is consistent with the deeper 4.8 GHz map
presented by Bridle et al. (1994) who label the radio core component 
``E'' and report it to be 1.1 mJy.  This is essentially the same as 
the value we measure, 1.05 $\pm$0.16 mJy (Table 2).  There is therefore 
no evidence for significant variability during the 12 years between observations.

Figure 4 shows that the radio core spectrum has a normal flat-spectrum
from 1.4 -- 22 GHz, observed frame, or 3 -- 50 GHz rest-frame
($z=1.228$), with a flux density of $\sim$1 mJy.  The
non-detection at the lowest frequency used, 325 MHz,
leaves us with an uninteresting upper limit; there could be
absorption at this frequency but we have no evidence for this.
So, what is the significance of the null result for absorption in the 
radio core?

It was this weakness of the radio core at 5 GHz relative to the
thousand-times brighter radio lobes (0.83 Jy) that in part suggested 
that the core might be absorbed at this frequency\footnote{The Bridle 
et al. (1994) core detection at 5 GHz is the only such measurement in the 
literature at any radio frequency for 3CR~68.1.}.  An observed-frame
frequency of 5 GHz corresponds to 11 GHz in the rest-frame of 
3CR 68.1.  Using a radio spectral index of $-1$ to characterize
the lobe spectrum and a flat radio core spectrum then implies a K-corrected
core dominance parameter at 5 GHz rest-frame of log R = $-$3.3, where 
R = $S_{core}/S_{lobe}$ (Orr \& Browne 1982).  This value is 
extremely small, an order of magnitude smaller than that of other
very lobe-dominated quasars (e.g., Wills \& Browne 1986).
For reference, the nearby edge-on radio galaxy Cygnus A 
shows a ratio of core-to-lobe flux at 5 GHz of 0.5\%, or log R = $-2.3$.

The line of sight optical reddening toward 3CR~68.1 is on the order of
A$_V = 1.7$ mag (Simpson \& Rawlings 2000) or perhaps slightly smaller
depending on the host galaxy contribution (Brotherton et al. 1998).
Using an estimate of E(B-V) = 0.5 then corresponds to a gas column density 
of N$_H = 3 \times 10^{21}$ cm$^{-2}$, given the standard gas-to-dust ratio for 
the Galaxy.  A Small Magellanic Cloud (SMC)  extinction law is more consistent
with the apparent reddening toward 3CR~68.1, and the SMC has a larger
gas-to-dust ratio, implying a larger value N$_H = 3 \times 10^{22}$ cm$^{-2}$.
Gas-to-dust ratios toward quasars may surpass Galactic values by as much as 
two orders of magnitude (Maiolino et al. 2001) so an associated gas column 
density of N$_H \sim 10^{22}$ cm$^{-2}$\ is probably a rough order-of-magnitude 
lower limit.

If the obscuring, outflowing gas along the line of sight toward
3CR~68.1 is local to the environment of the nuclear activity as
is generally believed for warm absorbers, the material would be
photoionized with temperatures of T$_e \sim 10^4$\ K.  

In the radio regime, the free-free opacity is a function of 
frequency, temperature, and the emission measure. 
From Osterbrock (1989):
\begin{equation}
\tau_{ff} = 8.24 \times 10^{-2} T^{-1.35} \nu^{-2.1} E 
\end{equation}
\noindent
where $T$\ is the temperature in units of degrees Kelvin, 
$\nu$\ is in GHz, and $E$\ is the emission measure in cm$^{-6}$\ pc.
Free-free absorption is insignificant at high frequencies,
but becomes strong at frequencies lower than the turnover
frequency at which $\tau_{ff} = 1$.  This sharp turnover signifies
free-free absorption and indicates a column of ionized gas along the 
line of sight.  For the redshift of 3CR~68.1 and a photoionization temperature
of T$_e \sim 10^4$\ K, equation 1 can be
rewritten to estimate the turnover frequency:
\begin{equation}
\nu_{o} \approx 15(n_{e5}N_{H22})^{0.5} {\rm GHz} 
\end{equation}
\noindent
where $n_{e5}$ is the electron density in units of 10$^5$ cm$^{-3}$, and 
$N_{H22}$\ is the ionized hydrogen column in units of 10$^{22}$\ atoms 
cm$^{-2}$.  Given that the total hydrogen column density is on order
of 10$^{22}$ cm$^{-2}$ or larger, we might well have expected to see 
free-free absorption in the frequency range we have investigated here.
Lines of sight through an obscuring torus are similarly predicted
to show free-free absorption in the range of a few to 10 GHz rest-frame
(e.g., Krolik \& Lepp 1989; Neufeld, Maloney, \& Conger 1994).
Free-free absorption has been seen in some edge-on systems with
warm absorbers, for instance NGC 4151 (Pedlar et al. 1998), but
not other edge-on systems (e.g., Barvainis \& Lonsdale 1998).

There are several possible explanations as to why we do not see
free-free absorption in 3CR~68.1 despite the above argument.
First, it may be that the radio core comes from a larger region than 
the optical continuum and is not
significantly covered by the outflow.  Also possible is that the
outflow is very low ionization (Mg I $\lambda$2853 is present)
and that the column density of ionized gas is too low to cause
free-free absorption at these frequencies; this might suggest
that the outflow is not close to the continuum source.

So if the radio core is not absorbed, could the extremely low
core dominance arise from over-luminous radio lobes?  After all,
the radio lobes reflect a previous epoch of activity that may
not be representative of the current activity level in the core.
Furthermore, radio-lobe brightness does not solely depend upon
properties of the central engine, but also environmental factors.
This question can be addressed by looking at the value of the alternative
core dominance parameter, R$_V$, which normalizes the radio core
using the optical continuum (Wills \& Brotherton 1995).  Using a
dereddened absolute magnitude M$_B$ = $-$28.2 (for $H_o = 
50$\ km s$^{-1}$ Mpc$^{-1}$ and $q_o$ = 0; Brotherton et al. 1998)
and a radio core luminosity of $2 \times 10^{24}$\ W Hz$^{-1}$,
then log R$_V$ = log L$_{core}$ + M$_B/2.5 - 13.69$, or 
log R$_V$ = $-$0.7 in the case of 3CR~68.1.  Comparison with the 
distribution of log R$_V$ for representative bright quasars (Brotherton 1996) 
shows that once again the radio core of 3CR~68.1 is underluminous
for its optical luminosity by at least an order of magnitude.
We conclude that the radio core truly is intrinsically very faint
despite the high optical and radio lobe luminosities of 3CR~68.1.
Anomalously strong Doppler de-beaming could be responsible for
the radio core faintness given a faster-than-usual jet, for instance,
but the faintness at GHz frequencies is not the result of 
absorption from material associated with the absorbers present 
in other wavebands.  3CR~68.1 apparently represents some kind of natural 
extreme for radio core-to-lobe ratios in quasars.

\section{Conclusions}

While 3CR~68.1 clearly has dust and gas along the line of sight that
causes absorption at optical and X-ray wavelengths, there does not
appear to be any significant absorption toward the radio core at
GHz frequencies.  The radio core spectrum is consistent with a flat 
($\alpha = 0$) spectrum of approximately 1 mJy, less than 1/1000 as bright 
as the radio lobes at 5 GHz, rest-frame, with no evidence for 
free-free or other absorption at radio frequencies.  The case
of 3CR~68.1 apparently represents the intrinsic low extreme of
the radio core dominance parameter used to characterize orientation
in unified schemes.

\acknowledgments

We thank Tony Beasley for help constructing an effective observing
program, and an anonymous referee for suggestions that improved this 
paper.  The National Optical Astronomy Observatories are operated by the 
Association of Universities for Research in Astronomy, Inc., under cooperative 
agreement with the National Science Foundation.
The National Radio Astronomy Observatory is a facility of the National Science 
Foundation operated under cooperative agreement by Associated Universities, Inc.
This research has made use of the NASA/IPAC Extragalactic Database (NED) 
which is operated by the Jet Propulsion Laboratory, California Institute of 
Technology, under contract with the National Aeronautics and 
Space Administration.   The work by W. v. B. at the Institute of Geophysics and
Planetary Physics at Lawrence Livermore National Laboratory was performed under
the auspices of the US Department of Energy by University of California
Lawrence Livermore National Laboratory under contract W-7405-ENG-48.


\begin{deluxetable}{lcccccc}
\tablewidth{0pt}
\tablecaption{VLA A-Array Observation Log, May 24, 1998}
\tablehead{
 Object & 22 GHz & 15 GHz & 8.4 GHz & 4.8 GHz & 1.4 GHz & 0.3 GHz\\
 & (sec) & (sec) & (sec) & (sec) & (sec) & (sec) }
\startdata
0137+331 (3C 48) & 70  & 50  & 60   & 60  & 80  & 150   \\
0237+288        & 550 & 550 & 660  & 620 & 460 &730    \\
3CR 68.1        & 1250& 1500& 1060 & 1110& 1240&1190   \\
0319+415        & 60  & 60  & 70   & 70  & 90  &\nodata\\
0521+166        & 60  & 60  & 70   & 60  & 80  &80     \\
\enddata
\label{table1}
\end{deluxetable}

\clearpage

\begin{deluxetable}{cccc}
\tablewidth{0pt}
\tablecaption{Radio Core Fluxes for 3CR 68.1}
\tablehead{
\colhead{Band} & \colhead{Frequency} & \colhead{Flux} & \colhead{RMS}\\
 & (GHz) & (mJy) & (mJy) }
\startdata
P & 0.3  & $\leq 8$   & 13\\
L & 1.4  & $0.65\pm0.42$   & 0.29\\
C & 4.8  & $1.05\pm0.15$   & 0.11\\
X & 8.4  & $1.04\pm0.07$ & 0.09\\
U & 15 & $0.89\pm0.28$  & 0.11\\
K & 22 & $0.69\pm0.24$  & 0.09\\
\enddata
\label{table2}
\end{deluxetable}

\clearpage


\begin{figure}
\plottwo{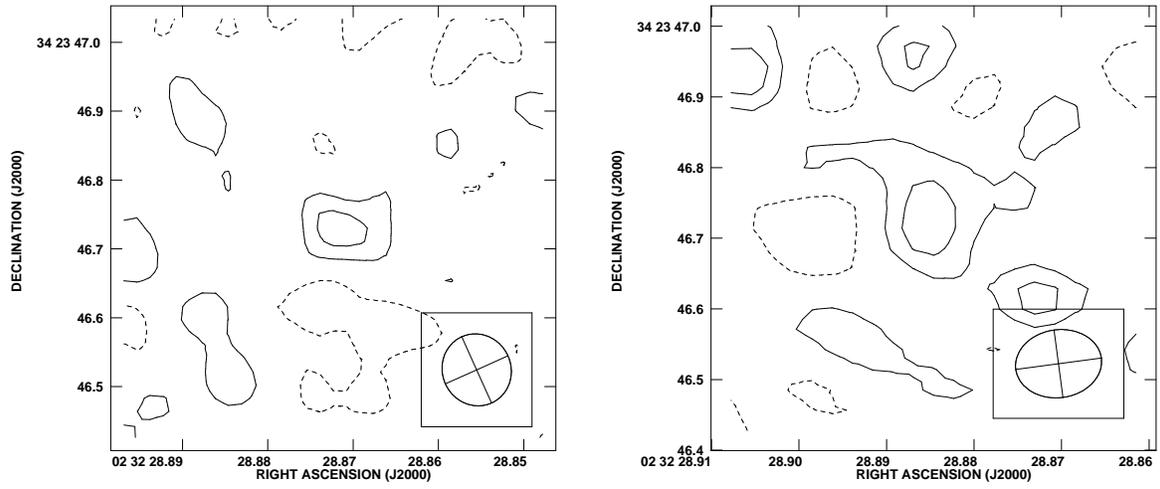}{Brotherton.fig1r.eps}
\caption{
Total intensity contour maps of the 3CR 68.1 core region at 22 GHz (K band, left)
and 15 GHz (U band, right).  Beams are plotted to show the resolution of the array.
Contour levels are drawn at 2.5, 5, 10, 20, etc. times the RMS flux as given
in Table 2.  The maps are centered on the core.
}
\label{fig1}
\end{figure}

\begin{figure}
\plottwo{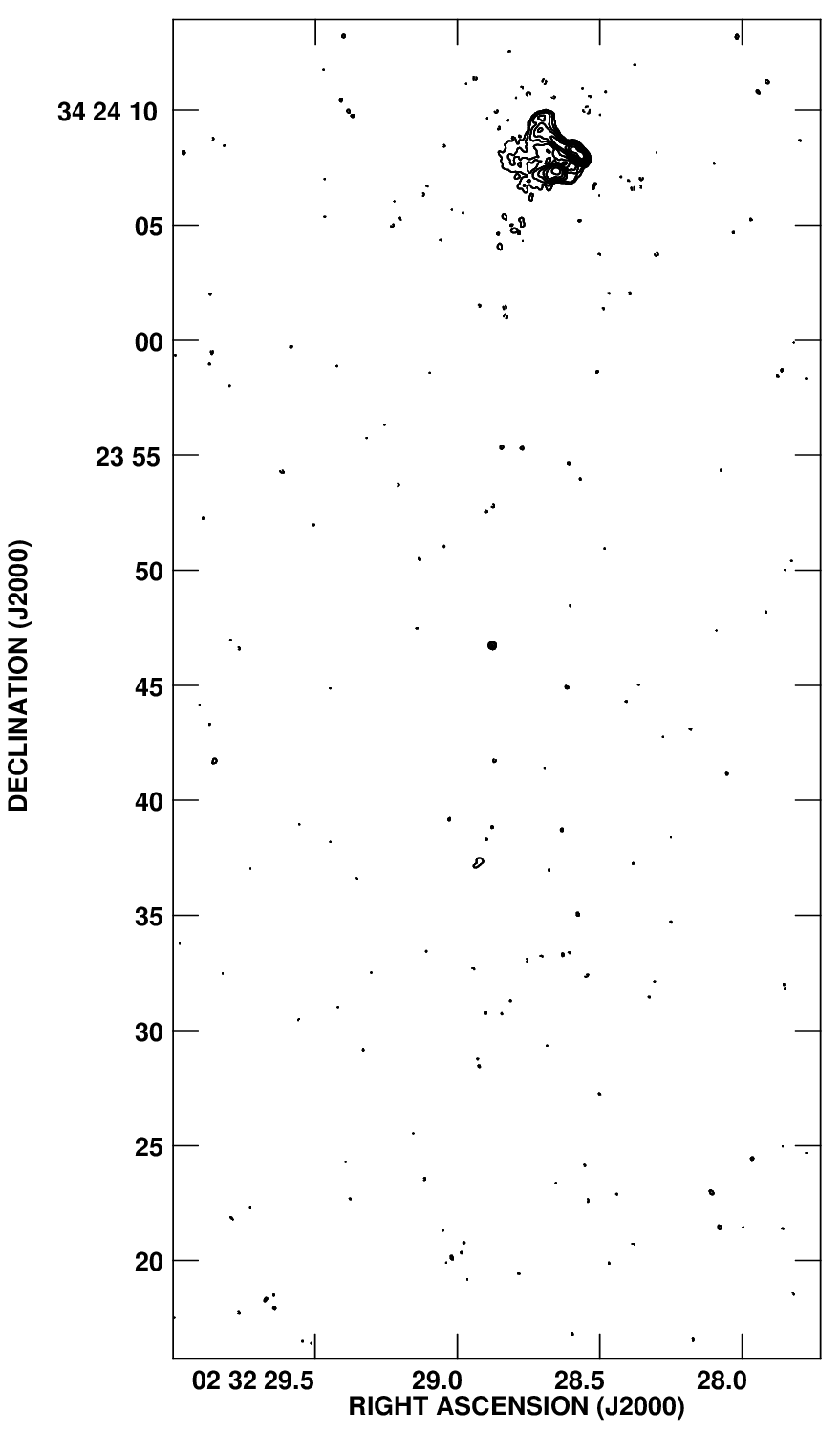}{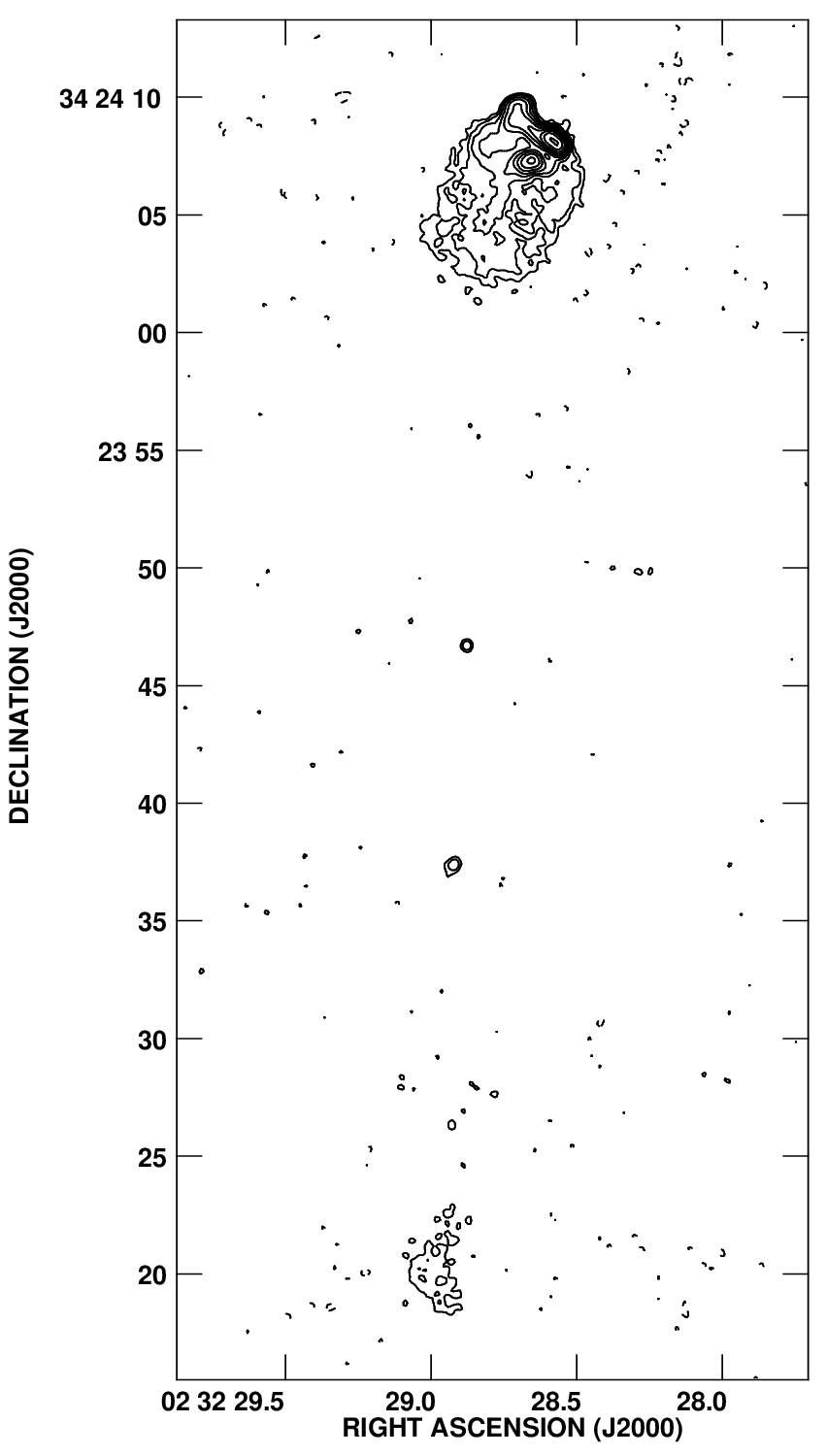}
\caption{
Total intensity contour maps of 3CR 68.1 at 8.4 GHz (X band, left) and 4.8 GHz
(C band, right).  Contour levels are drawn at 2.5, 5, 10, 20, etc. times the RMS flux 
as given in Table 2. 
}
\label{fig2}
\end{figure}

\begin{figure}
\plottwo{Brotherton.fig3l.eps}{Brotherton.fig3r.eps}
\caption{
Total intensity contour maps of 3CR 68.1 at 1.4 GHz, (L band, left) and 325 MHz 
(P band, right).  Contour levels are drawn at 2.5, 5, 10, 20, etc. times the RMS flux 
as given in Table 2.
}
\label{fig3}
\end{figure}

\begin{figure}
\plotone{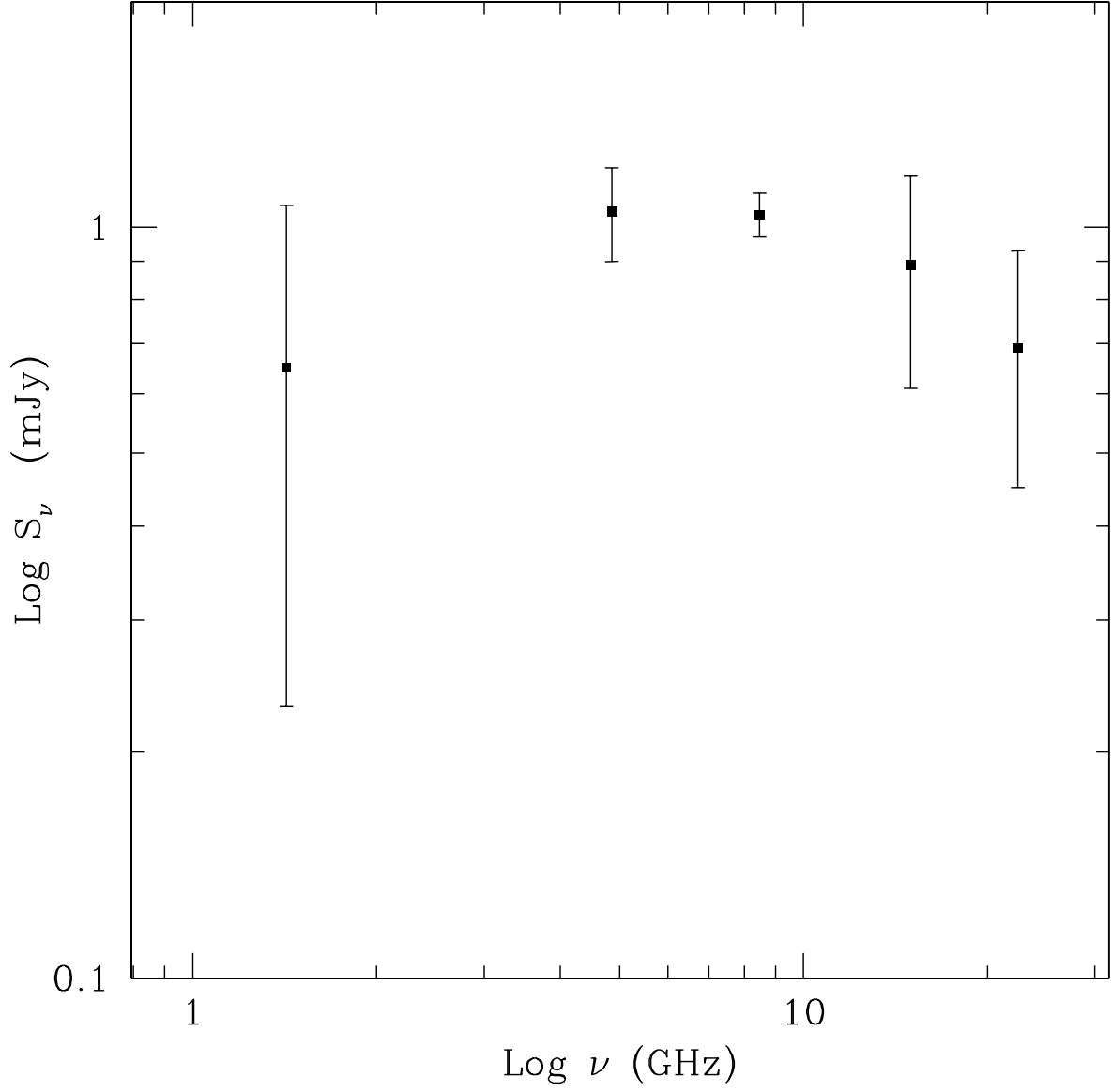}
\caption{The 3CR~68.1 core flux density $S_\nu$ vs. frequency $\nu$,
observed frame, with uncertainties described in the text and given
in Table 2.}
\label{fig4}
\end{figure}

\end{document}